# New Quantitative Study for Dissertations Repository System

Fahad H.Alshammari, Rami Alnaqeib, M.A.Zaidan, Ali K.Hmood, B.B.Zaidan, A.A.Zaidan

**Abstract —** In the age of technology, the information communication technology becomes very important especially in education field. Students must be allowed to learn anytime, anywhere and at their own place. The facility of library in the university should be developed. In this paper we are going to present new Quantitative Study for Dissertations Repository System and also recommend future application of the approach.

————————— ◆ —————————

## 1. INTRODUCTION

The objective of this deliverable is to introduce a number of system models for our Dissertations Repository System (DRS). It is important to develop our system according to the boundaries of the graphical representations described in our system models. The system models are to describe the context; sequence, class and data flow of our analysis and design processes. Based on the completed diagrams, once approved, we will use the models to develop a prototype of the DRS. We use the context diagram to show the environment of the system. This diagram represents an external perspective of the system. The architecture of the system represents a structural perspective of the system. We cover the DRS System Models in section 2.0 and the Architectural Design in section 3.0. We introduce each of the mentioned models in the Introduction of each system model diagram. Descriptions are provided for the important aspect of the DRS system. The system will allow students access to the library material through an online login system. In creating and hosting digital collections, the goal of this project is to:

- Provide visibility and access to dissertation from library to postgraduate students
- Create a model for the system of services and tools.
- Build a framework for the Dissertation Repository System (DRS).

## 2. METHODOLOGY

It is important for fallow the strategy of the research to choice the method that will be employed in the particular study. According to May (2001) "quantitative research aims to describe or explain the characteristics or opinions of a population through the use of a representative sample" [1], [6] and thus the author needs to use the quantitative research, however qualitative research is associated with participant's observation and interviewing and is preferable when the data of examination are complex and cannot be quantified [2], [6]. In this case the author will be in need for use qualitative research as well. Quantitative and qualitative methods involve different research strategies and the decision about which strategy to employ depends on the research situation [3], [6]. Quantitative research generally entails a more deductive approach to the relationship between theory and research in contrast with qualitative research that generally entails a more inductive approach. This means that the first one takes the theory as the starting point of investigation and functions to produce empirical evidence to test that theory.

### 2.1 Questionnaire

The use of a questionnaire is the most appropriate method to collect the quantitative data regarding the fact that is an inexpensive and fast way to gather data from a potentially large number of respondents [4],[6]. Therefore, comparing with the scheduled interviews, questionnaire is more convenient for responders because it offers them the freedom to complete it the time and at the speed they want [5] ,[6].

### 2.2 Demographic Features Respondents

The survey respondents were asked as to which age group they belonged to.

Table 1.
Breakdown in Terms of age, Nationality and Gender

| Age | | Nationality | | Gender | |
|---|---|---|---|---|---|
| 17 - 23 | 5 | Local | 20 | Male | 17 |
| 24 - 28 | 20 | International | 15 | Female | 18 |
| >28 | 10 | | | | |

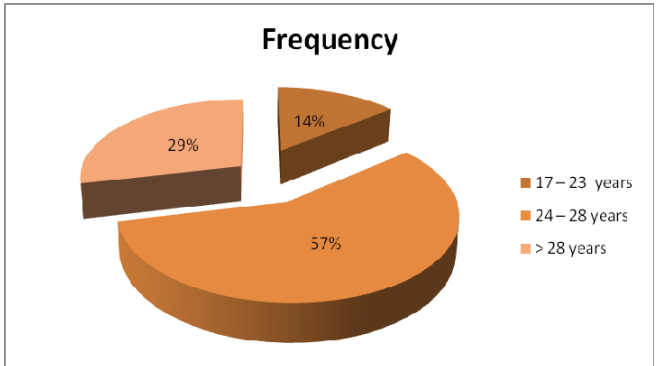

Fig 1. Frequency depend on the age

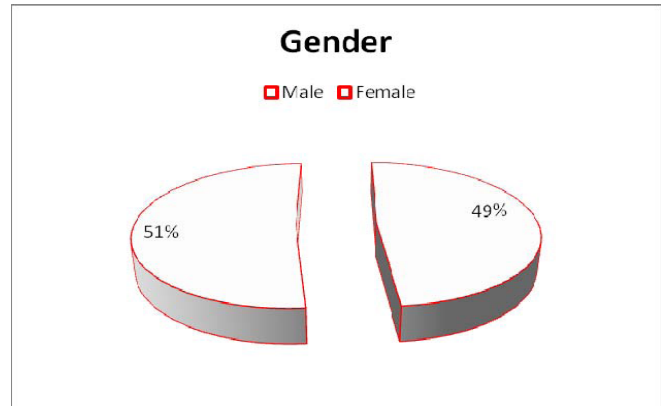

Fig 3. The Ratio of Male Respondents to Female

As depicted in table 1 and figure 1, the responses revealed was 14% of the respondents are from the 17 – 23 category while 57 % the respondents are from the 24 – 28 age category. In the more than 28 years of age category, 29 % of the respondents fall into this category.

Figures 2-5 of the respondents are international users and this indicates that quit big number of the respondents in the Kuala Lumpur area are from overseas. Figure 3 reflects that eighteen out of thirty five respondents are female users reflecting that the majority of the users are Female.

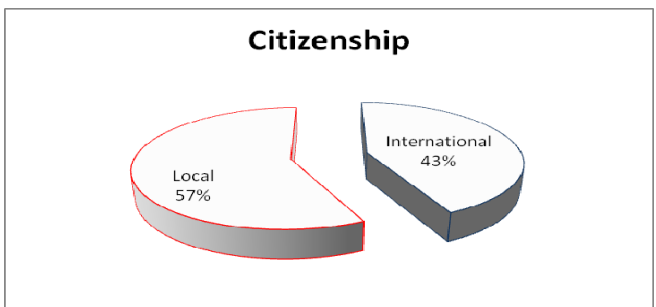

Fig 2. Ratio of local People in Relation to International People

For the question one which is, how long you have been using the internet, as it shown in figure 4, the 43% of respondents have used the Internet between one to five years. 50% have used it more than five years twenty five. Only 7% of the participants have experience less than one year.

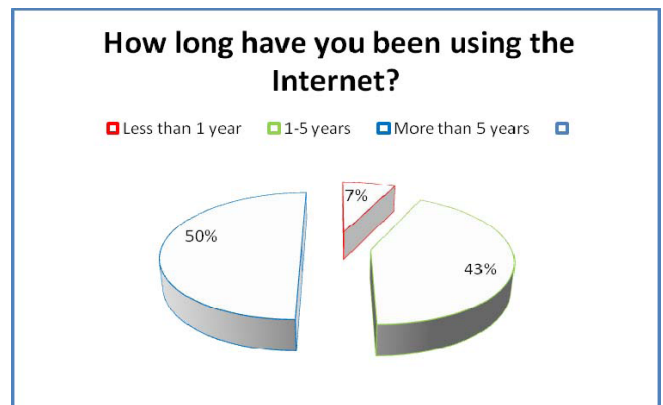

Fig 4. Internet Experience

In this question the author would like to make quantitative evaluation to the electronic forms at the faculty of computer science and information technology/ University Malaya and how far the people are comfortable with the current system, the respond shows the people mostly don't like or feel offended from the electronic forms over there. 48% ranked one, 26% ranked two, 17% ranked three, 3% ranked four, and 6% ranked five. Refer to figure 5.

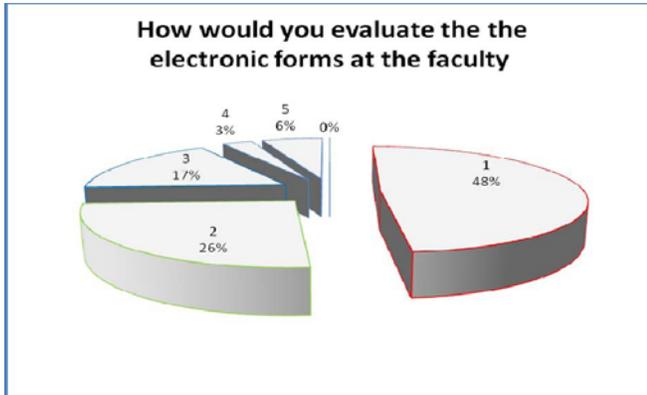

Fig 5. Evaluate the Electronic Forms

Next figure, figure 6 shows the ratio of the people who has been interrupted by the electronic forms at the faculty, as it has been clearly shown below, 80% from the respond said yes, we are upset from FSKTM form, while 14% said no we are ok with the system. Only 16% said we don't know [6].

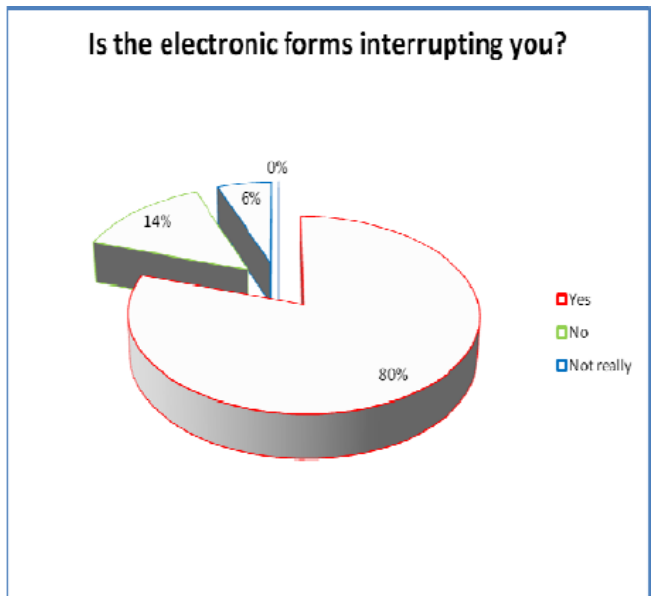

Fig 6. Electronic Form Interrupting

## 3. SYSTEM MODUEL

System modeling helps the analyst to understand the functionality of the system and models are used to communicate with customers.

### 3.1 Class Diagram

A class diagram is a type of static structure diagram that describes the structure of a system by showing the system's classes, their attributes, and the relationships between the classes.

Class diagrams are used for a wide variety of purposes, including both conceptual/domain modelling and detailed design modelling.

The most important usage of this diagram is to state all the classes and functions that they can do in one diagram and illustrate relationships between these classes, which will help you to come up with the basic data flow diagram. You can see which classes are related to each other and what methods they use in this kind of relationship. See Figure 1.7

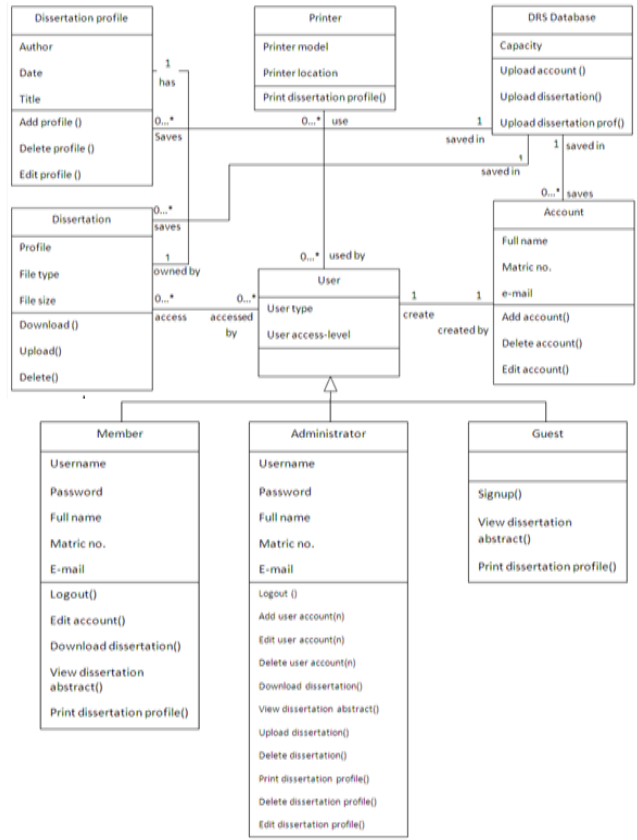

Fig .7 Class Diagram

### 3.2 Use case Diagram

A use case is a sequence of actions that provide a measurable value to an actor.

- Use Case Description:

  i. Use case description for login.

    Table 2:

Use Case Description for Login

| Use Case | Description |
|---|---|
| Login | This function appears on the first page and it is used to access the system. |
| Actor | Member |
| Precondition | Member must have a valid user ID and password. |
| Flow of events | 1. Member enters user ID and password in login form.<br>2. System checks the entered data.<br>3. If the data is valid, the Main Menu appears. |
| Post condition | Member can now access all the services provided by the system |

ii. Use case description for upload dissertation.

Table 3:
Upload Dissertation.

| Use Case | Description |
|---|---|
| Upload dissertation | Administrator uploads dissertation information to store it in repository. |
| Actor | Administrator |
| Precondition | The user must log in as administrator. |
| Flow of events | 1.Administrator enters user ID and password in login<br>2.Admin clicks "Upload" button in Admin window.<br>3.Upload form will come up.<br>4.Admin enters required data about dissertation such as: author, date, abstract.<br>5.Admin clicks upload button on the page.<br>6.Admin clicks close button to go back to Main Menu. |
| Post condition | The dissertation uploads successful and stored in repository. |

iii. Use Case Description for View Dissertation Abstract.

Table 4:
View Dissertation Abstract.

| Use Case | Description |
|---|---|
| View dissertation abstract | This function enables member to see abstract of the search result. |
| Actor | Member |
| Precondition | The member must be logged in.<br>Member press search. |
| Flow of events | 1.Member clicks the search button.<br>2.System displays the "Search" window with search form.<br>3.Member enters keywords.<br>4.System displays the results.<br>5.Member clicks the view button in front of each result.<br>6.An abstract info will appear.<br>7.Member clicks close button to close this new page. |
| Post condition | Member View dissertation abstract |

iv. Use case description for Delete user profile.

Table 5:
Delete user profile

| Use Case | Description |
|---|---|
| Delete user profile | Administrator can delete user profile. |
| Actor | Admin |
| Precondition | Admin must log in. |
| Flow of events | 1.System displays the "Admin Module" window<br>2.Admin search for user profile to be deleted.<br>3.System displays the user's record details.<br>4.Admin selects "Delete User".<br>5.System confirms deletion of record. |
| Post condition | User profile deleted successfully |

v. Use case description for Signup

Table 6:
Description for Signup

| Use Case | Description |
|---|---|
| Signup | When a new student registers in the faculty and gain matrix number can then sign up in the system and access to system. |
| Actor | Guest |
| Precondition | New user enters particulars. |
| Flow of events | 1.Guest enters matrix number, preferred username and password, personal information and a valid e-mail address. 2.System saves the profile. |
| Post condition | The guest can enter the system as member. |

## 3.3  Data Flow Diagram

Data flow diagrams illustrate how data is processed by a system in terms of inputs and outputs.

i.    Data Flow Diagram for Guest.

When the guest likes to register to be a member, he/she should first fill registration form. The information then validate, if it's valid he/she will be added to the members database. If the information is not valid he/she will be asked to fill another registration form. The guest can also search for a dissertation, if the dissertation found, he/she will see an abstract of the dissertation.

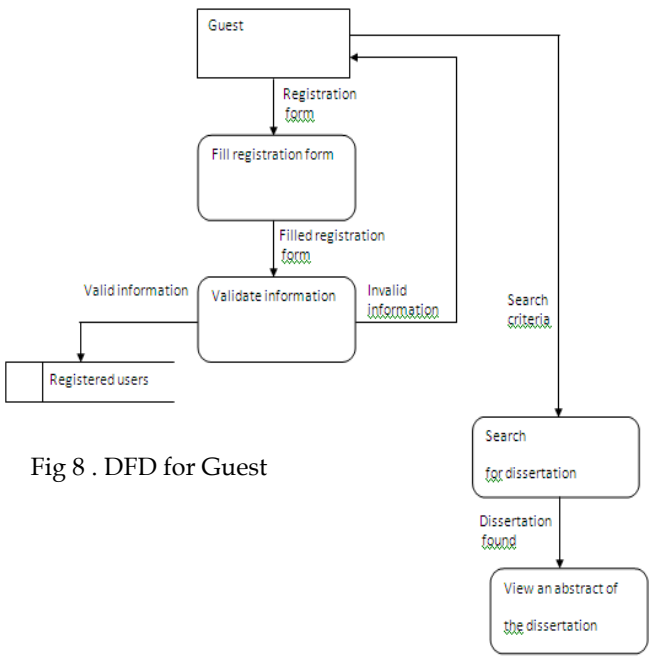

Fig 8 . DFD for Guest

ii.    Data Flow Diagram for Member

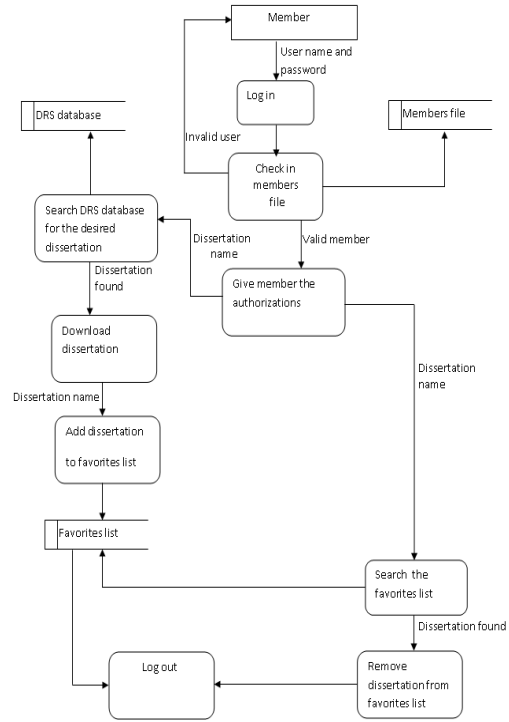

Fig 9.  DFD for Member

iii.    Data flow diagram for administrator

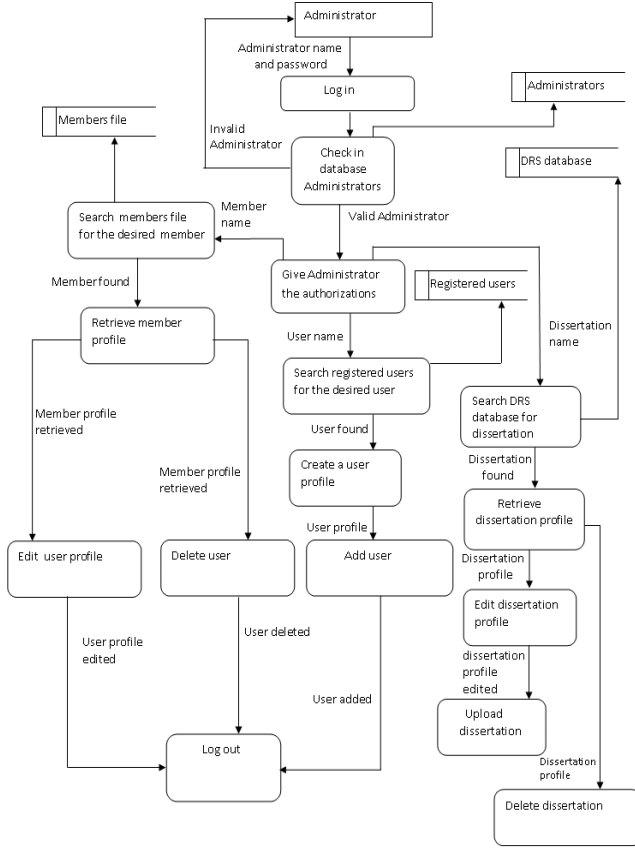

Fig 10. DFD for Administrator

## 3.4 Sequence Diagram

The sequence diagram (SD) specifies the time and control aspects of a system. Typically, SD is used to analyze only the more complex business events. Events are actions between the objects in the project. They can also transmit data.

A scenario is the sequence of events during one execution of a program. A scenario can include all the events or only the events sent to or received by certain objects in the system.

A sequence diagram shows, as parallel vertical lines, different processes or objects that live simultaneously, and, as horizontal arrows, the messages exchanged between them, in the order in which they occur. This allows the specification of simple runtime scenarios in a graphical manner, so that it would be easier for the user to follow the sequence of activities and flow of data.

i. Sequence Diagram of "Login" Use Case.

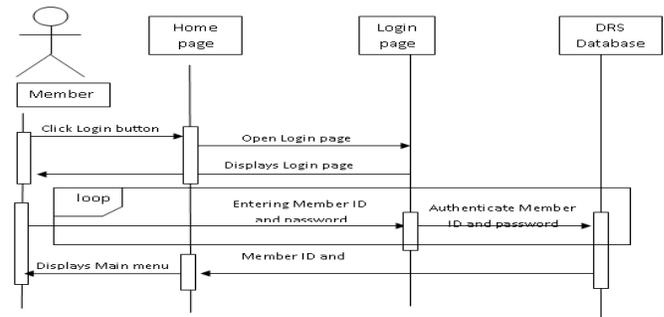

Fig 11. Sequence Diagram of "Login"

Brief Description:
This use case prompts the User to enter their username and password to be identified by the system when he or she wants to access the system.

Initial Step-By-Step Description:
Before this use case can be initialized, the user must be a valid user. User must have existing record by registering an account with the system.
1. Member enters his or her username and password in login form.
2. System validates username and password against DRS database.
3. If the information is correct, the Main Menu appears.

ii. Sequence Diagram of "Signup" Use Case

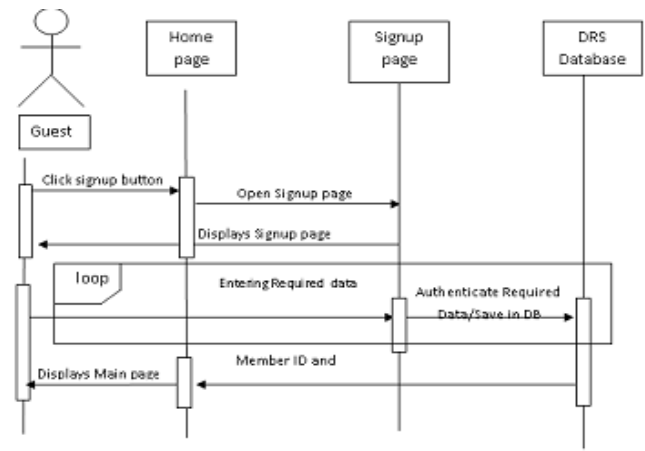

Fig 12 . Sequence Diagram of "Signup".

Brief Description:
This use case prompts the Guest to enter his/her matric number, username and password, some personal details and e-mail address to be saved in DRS Database for further use of the system. The DRS Database has a table of valid matric numbers which the Administrator entered them, so for a new member to signup in the system, first of all he/she should enter a valid matric number and after authentication of that, a valid username should be entered. So both matric number and username need to be verified.

Initial Step-By-Step Description:
To use this use case the user should be new in this system. To sign up he/she has to do the followings:
1. Guest enters his or her matric number, preferred username and password, personal information and a valid e-mail address.
2. System checks the entered information.
3. If the information is correct, the Main Menu appears.

iii. Sequence Diagram of "View dissertation abstract" Use Case

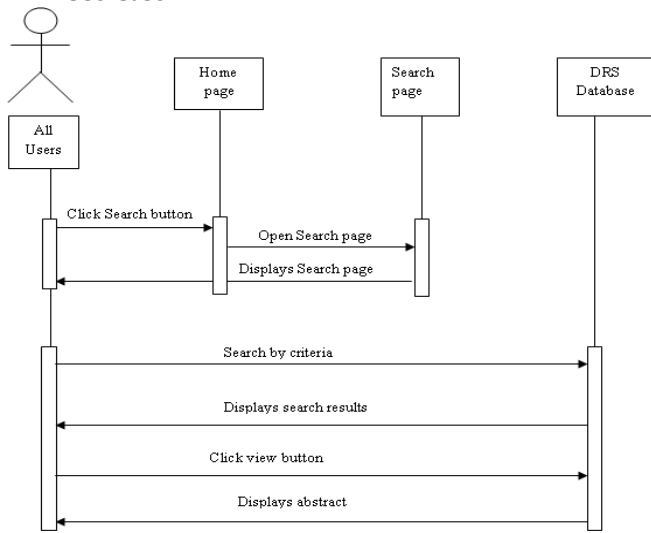

Fig 13 Sequence Diagram of "View dissertation abstract".

Brief Description:
This use case is initiated by all kinds of users (members, admin and guests) to view dissertations abstracts. The initial requirement for this use case is doing the search and having the results.

Initial Step-By-Step Description:
There is no initial use for the user to login for using this use case.
1. User clicks the search button.
2. System displays the "Search" window which consists search form.
3. User enters keywords in criteria.
4. System displays the results.
5. User clicks the view button in front of each result that he/she wants.

6. Another page appears consists the abstract of that dissertation.
7. User clicks close button to close this new page.

iv. Sequence Diagram of "Delete user profile" Use Case

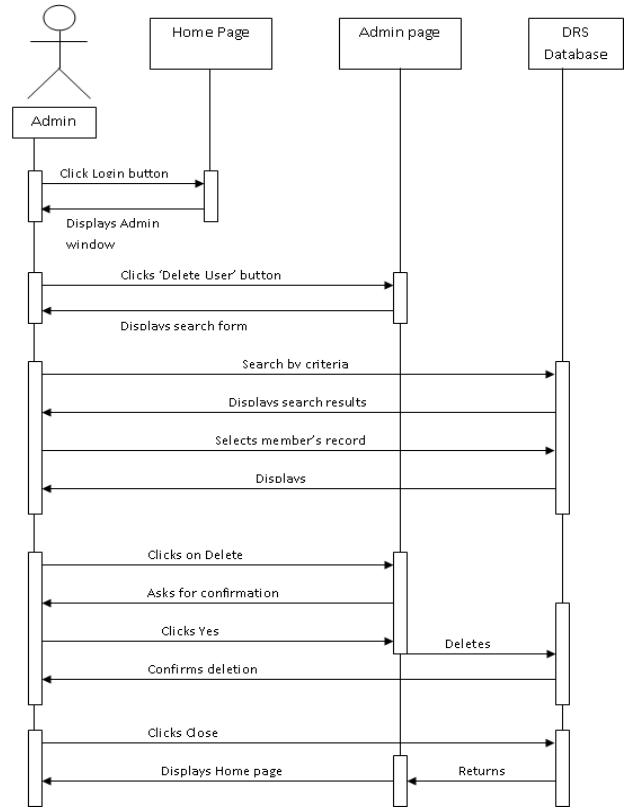

Fig 14. Sequence Diagram of "Delete user profile"

Brief Description:
This use case is initiated by the Admin to delete a user's record from the system after he / she has left the faculty.

Initial Step-By-Step Description:
Before this use case can be initialized, the user must have logged in as Admin and the record must exist in the system.
1. User login as Admin in the Main Menu.
2. System displays the "Admin Page" window.
3. Admin selects "Delete User" from the Admin Main Menu.
4. Admin keys in the keywords and searches for the record by clicking on a Search button.
5. Admin selects the user's record to be deleted from the search list.
6. System displays the user's record details.
7. Admin clicks "Delete".
8. Admin verifies choice to delete record.

9. System confirms deletion of record.
10. Admin clicks "Close" and system returns to the Membership Main Menu.

v.  Sequence Diagram of "Logout" Use Case

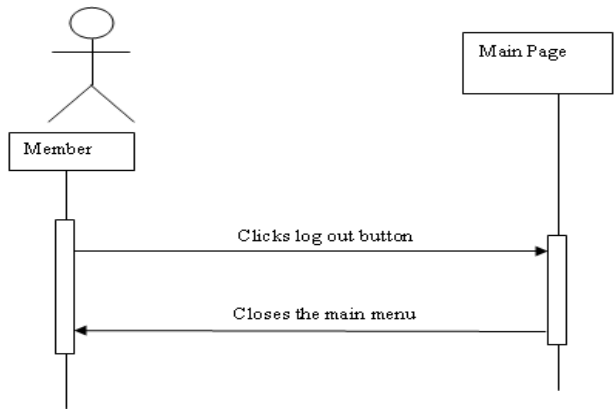

Fig 15. Sequence Diagram of "Logout" Use Case

Brief Description:
This use case logs out the member from the system.

Initial Step-By-Step Description:
Before this use case can be initialized, the member must be logged in as a user.
1. Member is at the Main Menu.
2. Member clicks on the Logout button.
3. System ends the session and closes the Main Menu.

## 4. ARCHITECTURAL DESIGN

Fully detailed systems architecture resolves into software and hardware systems. Desirable systems properties such as scalability and extensibility can be taken into account at the systems architecture level. The systems architecture is rationalized relative to the operational and technical architectures.

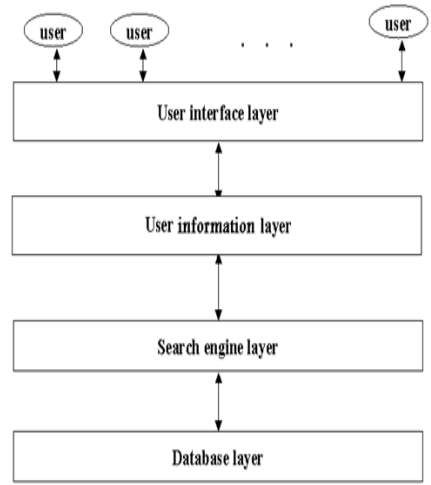

Fig 16. Architectural Design

Hardware component is the servers and the software component is the Web browser

- The suggest architecture is server and client architecture
- The link between them is the web browser
- We have database and it connect to the server
- This server will Provides two types of services related to the dissertations
   1. Everyone will be able to search among dissertations using different fields such as author, date, topic, etc.
   2. Valid users will be able to download the whole dissertation after finding it.
- Upload dissertations which related to the administrator
- Registration process and its classify to two type :
   - Register the dissertations
   - Register the new student (Sign up).
- Administrators will have the full access to the system to manage dissertations and users, such as add, delete, edit, etc.
- All kinds of users will be able to view any dissertation's abstract after finding it, whether being a member or not.

## 5. CONCLUSION

Design for the dissertation repository system (DRS) was involved on deciding the data structures, classes, algorithms and proceduress to be used. It is a part of design and a kind of design. System design is focussed on system model of the DRS and is emphasised on sequence diagram, class diagram, data flow diagram and arthitectural design. First of all, we start working on in the current release and updated our static model (class diagrams of the DRS). The purpose of the sequence diagrams is to drive the design. But we advocate drawing our sequence diagrams of our DRS in a minimal and a quite specific format. Then we continued with data flow diagram to show the sequence or flow of the DRS. The user group is made of user type and user access-level, then categorized to three groups; member, admin and guest. Both member and admin group is made up of username, password, full name, matric number and e-mail. The admin has the authority to add, edit and delete the user account and dissertation profile. Guest only can sign up, view dissertation abstract and print dissertationprofile. User is created by account class, which includes full name, matric number and email of students. Dissertation profile (author, date, title & add,edit and delete profile) has dissertation class ( profile, file type, file size & download, upload and delete dissertation) is accessed by user. Dissertation profile owned by database group, which has capacity, upload account, dissertation and dissertation profile. Database has the accounts of user or students.  At last we worked on the architectural design of the DRS. Fully detailed systems architecture is resolved into software and hardware systems. The systems architecture is rationalized relative to the operational and technical architectures. We created a direct link between each of them/ use case, its context diagram, sequence diagram, class diagram, data flow diagrams and the architectural design. This had built/ created the complete system design of our DRS. Hereby, we had illustrated on how to perform effective design reviews in our DRS.

**ACKNOWLEDGMENT**

This research has been funded in part from multimedia University; the author would like to acknowledge the entire worker in this project, and the people who support in any way, the author would like to thanks his friends who has support in many ways

**REFERENCES**

[1] Tim May, 2001, Social Research: issues, methods and process, 3rd edition, Open University Press, Buckingham.
[2] T. D. J. Chappel,(1999), "Understanding HumanGoods", ISBN:0748610294.
[3] Yin, (1994) ," Tourism strategy making: Insights to the events tourism domain ", Volume 29, Issue 2, April 2008, , School of Advertising, Marketing and Public Relations, Faculty of Business, Queensland University of Technology, GPO Box 2434, Brisbane 4001, Queensland, Australia,Pages 252-262
[4] Oppenheim, (1966)," Questionnaire design and attitude measurement ", / A. N. London : Heinemmann, Material Impreso.
[5] Alan Bryman (2001) , Social Research Methods (Paperback). by "The chief aim ...Paperback: 560 pages; Publisher: Oxford University Press, USA .
[6] Ibrahim A.S.Muhamadi, S.Raviraja, B.B Zaidan, A.A Zaidan, M.A Zaidan, Chengetai Mapundu (2009), "New Quantitative Study for the Student Record Retrieval System". IJCSNS International Journal of Computer Science and Network Security, VOL.9 No.8.